# Implementation of fast ICA using memristor crossbar arrays for blind image source separations

*Pavan Kumar Reddy Boppidi[1], Victor Jeffry Louis[1], Arvind Subramaniam[1], Rajesh K. Tripathy[1], Souri Banerjee[2], Souvik Kundu[1]* ✉

[1]*Department of Electrical & Electronics, Engineering, Birla Institute of Technology and Science (BITS) Pilani, Hyderabad Campus, Hyderabad-500078, India*
[2]*Department of Physics, Birla Institute of Technology and Science (BITS) Pilani, Hyderabad Campus, Hyderabad-500078, India*
✉ *E-mail: souvikelt@gmail.com*

**Abstract:** Independent component analysis (ICA) is an unsupervised learning approach for computing the independent components (ICs) from the multivariate signals or data matrix. The ICs are evaluated based on the multiplication of the weight matrix with the multivariate data matrix. This study proposes a novel Pt/Cu:ZnO/Nb:STO memristor crossbar array for the implementation of both ACY ICA and Fast ICA for blind source separation. The data input was applied in the form of pulse width modulated voltages to the crossbar array and the weight of the implemented neural network is stored in the memristor. The output charges from the memristor columns are used to calculate the weight update, which is executed through the voltages kept higher than the memristor Set/Reset voltages (±1.30 V). In order to demonstrate its potential application, the proposed memristor crossbar arrays based fast ICA architecture is employed for image source separation problem. The experimental results demonstrate that the proposed approach is very effective to separate image sources, and also the contrast of the images are improved with an improvement factor in terms of percentage of structural similarity as 67.27% when compared with the software-based implementation of conventional ACY ICA and Fast ICA algorithms.

## 1 Introduction

In this digital era, the internet of things (IoT) has become a major attraction for operating machines wirelessly. Recently, there has been renewed interest in solving 'big data' which is also associated with IoT. In this regard, a system is required which would analyse, manage or detect different events, predict outputs and process data within a tolerable elapsed time [1]. Independent component analysis (ICA) is an unsupervised learning algorithm in which the multivariate data matrix is represented as the statistically independent variables, which became popular for blind source separation and other IoT-based applications [1]. Over the years, it has also found application in a wide array of fields such as signal processing, communication, speech recognition and medical science [2, 3]. A well-known example of its application has further been extended into de-noising images, extracting features and electrocardiogram (ECG) signals [1]. Neural network-based approaches have been proposed for extracting independent components (ICs) from the mixed signal [4–7]. For this application, the neural network was mostly implemented using CMOS device [8, 9]. However, CMOS is suffering from scaling, power and speed issues since the last few years which reached to the Von Neumann bottle neck. Therefore, the research took different directions where an alternative device is required which could ultimately replace CMOS device without compromising any performances. Memristors can be considered as a potential alternative to this effort, which is being used almost in every sector in electronics such as logic circuits, analogue–digital circuits, neuromorphic circuits and memory operations [10]. To implement neural networks, the memristor was found to be a great choice owing to its natural ability to perform matrix–vector multiplication (through varying its weight), which is considered as the fundamental operation for the neural network [11]. Moreover, memristor offers other advantages such as high retention, low power consumption, high speed switching, etc. [12, 13]. Research on memristors-based neural network was much celebrated in the area of sparse coding, PCA-based dimensionality reduction and image processing [14–16]. However, only two works have been attempted to develop memristor neural network based ICA. In this regard, Rák *et al.* [17] have investigated current controlled memristors to implement Rossenblatt's learning rule. In this work, multiple memristors have been used to implement one weight, which suffers from multiple areas and power consumption [1, 17]. In another work, Fouda *et al.* [1], implements error-gated Hebbian rule on filamentary $TiO_x$ memristors (which obey the conductance updated model) to retrieve Laplacian signals, but not the images. Their work also suffers from using a million programming pulses to program the weight of a memristor, which may drastically increase the power consumption and latency, ultimately hinder the overall performance of the networks. In addition, the proposed circuit may break down after implementing ICA for a few hundred times since the conduction mechanism in the conventional $TiO_2$ is based on filament model [18]. It is also important to mention that the abovementioned works have implemented the older versions of ICA, which are not as effective as entropy-based ICA, the latter is better in terms of mutual information, computationally efficient and requires less memory as compared to other ICAs [19, 20]. Moreover, the conventional ICA is not flexible; concede inadequate separation performance [20]. Hence, an alternative ICA approach, as well as, different material-based memristor (where conduction mechanism is non-filamentary based) should be investigated which may overcome the problems mentioned above and find out some important applications which include separation of images, medical ECG data, de-noising images etc. The usual approach to implement ICA is field-programmable gate array (FPGA), which is not only slower but consumes more power [21–23]. Application-specific integrated circuits (ASICs) were also considered as an alternative approach [24, 25]. However, both the FPGA and ASICs are based on CMOS devices and limited by the Von-Neumann bottleneck. Moreover, they are capable of implementing only the given specific algorithm and will not be able to implement other neural networks, which include memristor crossbar. Unfortunately, apart from only two literature works [1, 17], no such work is available in literature which deals with memristor neural network based ICA for electronic applications.

Recently, to mitigate the problems associated with conventional ICA, many improved ICA algorithms were proposed [26]. However, due to the presence of fast convergence property, the fast fixed-point algorithm, also known as Fast ICA was found to be the most promising one for blind source separation applications [26]. Unfortunately, till now, neither any memristor-based Fast ICA implementation has been attempted nor any of its application was demonstrated. Therefore, not much work has been performed in elucidating these and this gap retains scope for further investigation.

In this paper, to implement ICA, we have put our efforts to construct a standard neural network proposed by Amari (ACY) *et al.* [6] through utilising a voltage-controlled memristor crossbar arrays by utilising memristor as a synapse. Program of different weights, weight multiplication by the input signal and non-volatile weight storage are considered as the critical challenges in the execution of synaptic operation [27]. In this work, all these were accomplished by carefully utilising memristors. Among other neural network-based algorithms, ACY was preferred due to its relatively simple weight update rule and most optimal in terms of mutual information [26]. For the memristor, a new alternative non-filamentary Cu (5%) doped ZnO (Cu:ZnO) based Pt/Cu:ZnO/Nb:STO memristor was utilised. The detailed working mechanism of Pt/Cu:ZnO/Nb:STO memristors and their suitable electrical performances such low Set/Reset voltages, high retention, fast switching and non-filamentary conduction can be found in our earlier work [28]. The main contribution of this paper is to introduce an innovative memristor-based advanced neural network proposed by Hyvarinen [19], which is also known as Fast ICA, and this non-conventional attempt has been initiated for the first time to the best of our knowledge, though it possesses several advantages over conventional ICA as already discussed in the previous paragraph. Finally, in order to demonstrate an electronic image separation application, the robustness of the framework has been evaluated by employing it to retrieve individual faces from mixed images obtained from face-recognition data set [29]. Apart from facial image separation, this work can be used for separation of any images in general. Image separation has applications such as multiple fingerprint separation, image de-noising, handwritten digits/letters separation [2]. Further, the outcomes have been compared with the conventional ICA and software algorithm to understand the efficacy of Fast ICA for blind source separation application. All the important parameters such as the mean square error (MSE), the structural similarity (SSIM), the peak signal to noise ratio (PSNR) and gradient-based image quality assessment (GSM) have been calculated and the improvement in terms of the percentage of these measures are also evaluated and compared with the conventional software-based implementation.

This manuscript was organised as follows: in Section 2, we have briefed ACY ICA and Fast ICA. In Section 3, we have shown the role of Cu: ZnO memristor as a synapse and finally in Section 4, we have described the working mechanism of the proposed network and compared the output performances with the traditional software-based one.

## 2 Theory of ICA

The source signal $x(t)$, which is fed to the ICA algorithm is a linear combination of statistically ICs as given in [6]

$$x(t) = As(t) = \sum_{i=1}^{m} s_i(t) a_i \quad (1)$$

In order to obtain each IC in the mixed signal, $x(t)$, a number of neural network-based approaches have been proposed over the years [4–7]. The estimate of each IC, $y(t)$, is in the form of a neural network with $W$ as the weight matrix that needs to be trained [6].

$$y(t) = Wx(t) \quad (2)$$

In this paper, efforts were devoted to develop the neural network as proposed by Amari *et al* [6], owing to its relatively simple weight update rule as compared to other ICAs, as shown below

$$\Delta W = \mu_k [I - g(y) y^T] W \quad (3)$$

where $\mu_k$ is the learning rate and $g(y)$ is the non-linear activation function given by (4) in which $y(t)$ used is the same as in (2) as shown in [6]

$$g(y) = \frac{3}{4} y^{11} + \frac{25}{4} y^9 - \frac{14}{3} y^7 - \frac{47}{4} y^5 + \frac{29}{4} y^3 \quad (4)$$

However, the above given algorithm is not as efficient as Fast ICA in retrieving mutually ICs [26]. Fast ICA requires whitening of data. Whitening is done by the method prescribed in [2] and whitened data is represented by $v$. The weight update rule for which is shown in (5) and (6) as given in [19]

$$w_{i+1} = E\{vg(w_i^T v)\} - E\{g'(w_i^T v)\} w_i \quad (5)$$

$$w = \frac{w_{i+1}}{\| w_{i+1} \|} \quad (6)$$

where $w_i$ is the weight in the ith iteration and $w_{i+1}$ is weight in the $(i + 1)$th iteration. v is the whitened data as mentioned above and $E$ is the expectation which in our work is given by the sample means. The activation function is given by $g$ which is shown in (7) and $g'$ is its derivative which is given by (8) [19].

$$g(y) = y \exp\left(\frac{-y^2}{2}\right) \quad (7)$$

$$g'(y) = (1 - y^2) \exp\left(\frac{-y^2}{2}\right) \quad (8)$$

## 3 Memristor as a synapse

Recently, memristors have been found to be useful in emulating important features of biological synapses owing to their nanoscale dimensions, their ability to store multiple bits of information per element and the low energy required to write distinct states [30]. A Pt/Cu:ZnO/Nb:STO memristor was fabricated and the detailed fabrication procedures are provided in our earlier publication [28]. Since the resistance of a memristor varies with the amount of current passing through the memristor, it can store the weights of the neural network and update them using simple voltage pulses [31].

## 4 Methodologies

A circuit (shown in Fig. 1) was proposed to record its memristance and weight variation as a function of time. In the circuit depicted in Fig. 1, the pulses of amplitude −1.30 V were applied for 2 ns onto the top Pt electrode of the Cu:ZnO memristor, whereas 80 mV was enforced for 23 ns to perform the read operation. This process was repeated for every 500 ns during this period, the applied negative voltage cannot further reduce the area indexed of polarised material, $X$ (not below zero), which causes the resistance to remain constant at $R_{on}$ (40 kΩ), where $R_{on}$ is the resistance at low resistive state. Further for next 500 ns, the positive 1.50 V pulses were employed for 2 ns and the read operation was executed. To implement this, Cadence Virtuoso® 6.1.1-64b and Matlab were used. In this duration, an increase in $X$ caused the resistance to increase till 1000 ns. This sequence was repeated till 2600 ns. From 1000 to 1500 ns, the negative pulses reduced the value for $X$ and consequently the memristance. The memresistance of the memristor and the area index of the polarised material with time are shown in Fig. 2. From Fig. 2, one could notice that the pattern from 500 to 1500 ns repeats itself from 1500 to 2500 ns. The reason for choosing these time durations is to demonstrate the memristive variations (shown in Fig. 2).

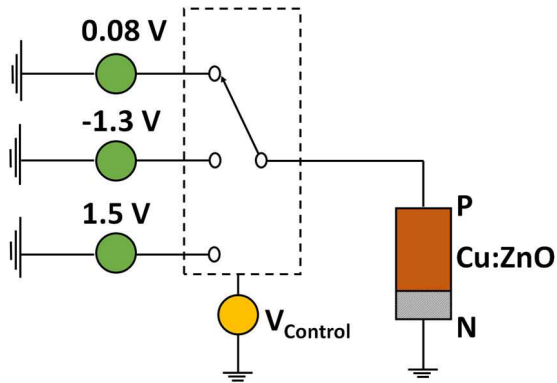

**Fig. 1** *1.5 and −1.3 V are used for writing a new resistance state in the Cu:ZnO memristor while 0.08 V is used for reading. Switching between these three voltages is interacted by $V_{control}$, which controls the switch*

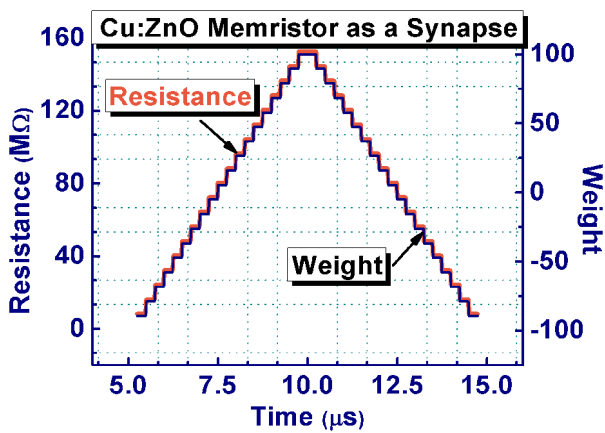

**Fig. 2** *Resistance and weight are plotted as a function of time. This graph is obtained using the circuit in Fig. 1. As the weight traces resistance, one can infer that resistance is proportional to weight*

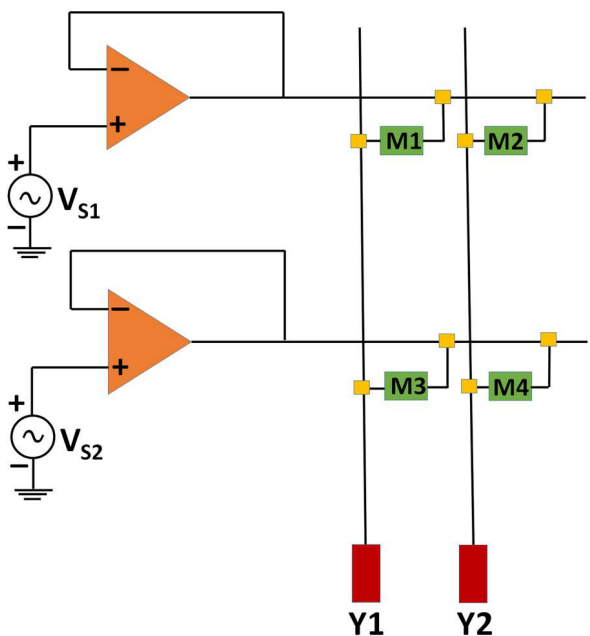

**Fig. 3** *Schematic representation of the Cu:ZnO memristive crossbar array. The inputs are fed in terms of voltage pulses through $V_{s1}$ and $V_{s2}$. The output charge is collected in Y1 and Y2*

The read period was kept greater than the write period to prevent the combining of two consecutive resistive states. From Fig. 2, it was understood that the memristance state is dependent on its previous state and applied voltage. In order to map the weight change in a memristor, the parameter $(X/D)$ was used, where $D$ is the active material thickness (in this case 50 nm). The area occupied by the polarised material is $X \times D$, and it is proportional to $(X/D)$ as $D$ is a constant. Hence, $(X/D)$ was considered owing to its range from 0 to 1 and reduced computational complexity. As weight varies from 100 to −100, (9) was utilised to map $(X/D)$ to the weight values. The weight values were measured along with a resistance change from 500 to 1500 ns and plotted as a function of time. Interestingly, it was found that the resistance can precisely trace the weight, which further ensures that a memristor can be used to map the neural network's weight.

As shown in (9), 100 is multiplied to scale the weights between −100 and 100 to match with the software algorithm weight. It is worth to mention that the initial weights can be negative [1]. In this case, the weights were also considered negative initially; however, the final output would be an image with finite pixel values, thus the weights would be changed to positive.

Here, $w_{ij}$ is the weight of one memristor. The weights of the algorithm are stored in the memristor itself due to having its non-volatile property [28].

$$w_{ij} = 100\left(\frac{2X}{D} - 1\right) \quad (9)$$

In order to realise matrix multiplication of the input data matrix with the weight matrix of the used neural network algorithm, a Cu:ZnO (memristor using VTEAM model, which mimic exactly the same output performances to the fabricated devices) based crossbar array has been developed (shown in Fig. 3). $V_{s1}$ and $V_{s2}$ are the voltage sources, which provide pulse width modulated voltage pulses for the inputs. Voltage follower amplifier was used to provide high input impedance. Initially, the charge at the output of crossbar architecture is obtained by utilising the relation given in (10) and is stored in charge collectors $Y1$ and $Y2$ as shown in Fig. 3.

$$Q_j = \sum_i v_i t_i / R_{ij} \quad (10)$$

Here, $v_i$ is the voltage applied to the ith row of the crossbar and $t_i$ is its corresponding time period. The magnitude of $t_i$ is proportional to the magnitude of the input (pixel value in this case). The magnitude of the corresponding voltage is 0.50 V which is less than the Set voltage of Cu:ZnO memristor. $R_{ij}$ is the resistance of the memristor at the ith row and jth column and is expressed in (11) [30–32] of VTEAM memristor model:

$$R_{ij} = (R_{off} - R_{on}) \times \frac{X}{D} + R_{on} \quad (11)$$

$R_{off}$ is the resistance in the high resistance state. As per VTEAM model, the change in $X$ with respect to time is given by

$$\frac{dX}{dt} = K_{off}\left(\frac{v}{v_{off}} - 1\right)^{\alpha_{off}} u(v - v_{off}) + K_{on}\left(\frac{v}{v_{on}} - 1\right)^{\alpha_{on}} u(v_{on} - v) \quad (12)$$

where $u$ is a unit step function of applied voltage v. $K_{off}$ and $K_{on}$ are the constants which control the transition from one state to another. Whereas, $\alpha_{off}$ and $\alpha_{on}$ are also the constants which were used to fit non-linearity [30, 31].

Equation (9) gives us the way in which weights were stored in terms of $X$ and (11) tells us how $X$ affects the resistance of a memristor. Fig. 4 shows us how the weight was stored in the form of resistance of the device. This data was observed for one memristor used in the Fast ICA implementation. Note that, the weight stabilises after the fourth cycle and this pattern is reflected in the resistance of the device. As the resistance closely follows the weight values, the weight has been successfully stored within the memristor. It is noteworthy to mention that in some memristive devices, the magnitude of the weight change was found to be

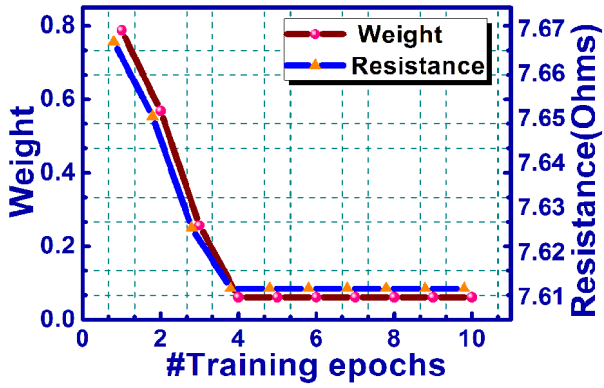

**Fig. 4** *This graph represents the resistance in one of the four Cu:ZnO memristors in the crossbar and also the weight value stored in it*

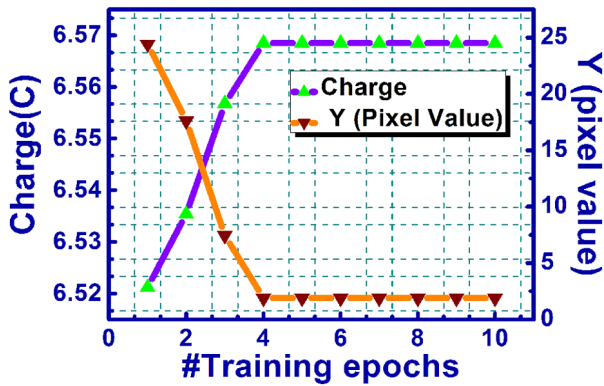

**Fig. 5** *This graph depicts the charge in one of the four Cu:ZnO memristors in the crossbar and also the pixel value obtained from the charge*

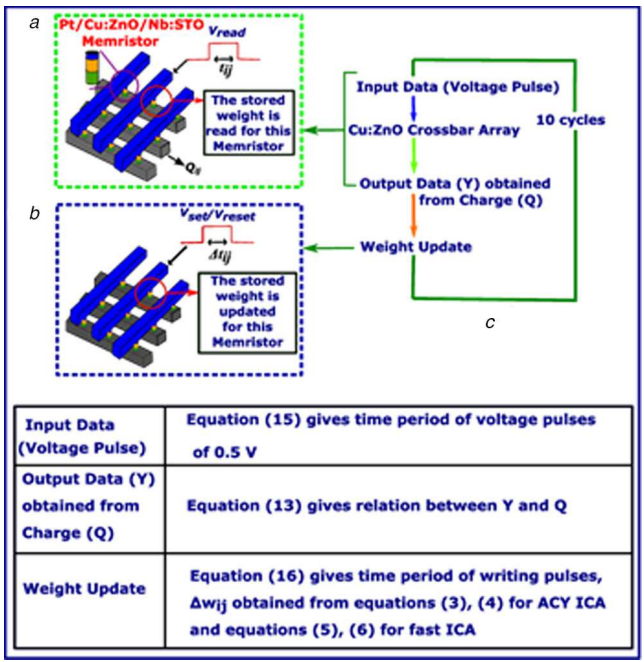

**Fig. 6** *Flowchart depicts the mechanism*
*(a)* Process for mapping input data to input voltage pulses in the crossbar, *(b)* Charge Q was used to calculate the weight update as per the algorithm and the weights were updated, *(c)* Proposed implementation flowchart higher for the initial cycles and gets lower for the final cycles [33, 34]. However, to implement such devices, the authors have used models different from ours (VTEAM in this case). Therefore, in the Cu:ZnO memristors, the weight change is dependent only on the applied voltage pulse and amplitude, not on the interval between pulses. The VTEAM model and Matlab were employed to order to construct the proposed network and observe the output characteristics.

## 5 Results and discussion

The component $y_j$ was obtained from $Q_j$ as per (13), which contains the two original input images pixels (where the ith feature is $c_i$) as given in (2)

$$y_j = \sum_i 100\left[2\left(\frac{(v_o t_i/(Q_{ij})) - R_{on}}{R_{off} - R_{on}}\right) - 1\right]c_i \quad (13)$$

By carefully choosing $y_j$, one could find out the output which is a product of the weights and the inputs as expected in a neural network algorithm (as per (2))

$$y_j = \sum_{i=1}^{n} g_{ij} c_i \quad (14)$$

Here, $g_{ij}$ is the weight and $c_i$ is proportional to the ith pixel. In this case, the $c_i$ would be equal to the pixel value of the input (mixed) image and the pixel was ranged from 0 to 256. The relation between $c_i$ and $t_i$ is given by

$$t_i = t_0 \times c_i \quad (15)$$

where $c_i$ is a dimensionless quantity proportional to the input and $t_0$ is the time constant of 100 μs. It is noteworthy to mention that, (13) is extremely essential to implement any neural network in the form of (2) on the crossbar array.

Equation (13) gives us the way to obtain $y_j$, the output which is the pixel value from the charge. From Fig. 5, we can see that $y_j$ reduces as charge increases. This is due to the inverse relationship as between both these quantities. Moreover, once the weight settles, the charge also remains the same. This, in turn, leads to the pixel value also remaining constant from the fourth cycle onwards, as seen in Fig. 5.

During the first phase, 0.50 V pulses were applied with pulse widths proportional to the corresponding pixel values in order to obtain the output $y_j$ as per (13). As 0.50 V is less than $v_{off}$, it does not alter the memristance. In the second phase, the weights of the memristors were changed as per (3). For positive results of the change in weight, $\Delta g_{ij}$, pulses of amplitude +2 V were applied, whereas for negative results, pulses −2 V were used in order to precisely update the memristive weights. The pulse widths ($\Delta t_{ij}$) for the above two cases were determined by (16).

$$\Delta t_{ij} = \frac{(((\Delta w_{ij}/100) + 1/(2))D)u(\Delta g_{ij})}{K_{off}((v/v_{off}) - 1)^{\alpha_{off}}} + \frac{(((\Delta w_{ij}/100) + 1/(2))D)u(-\Delta g_{ij})}{K_{on}((v/v_{on}) - 1)^{\alpha_{on}}} \quad (16)$$

It can be observed that the requisite voltage was applied to only one memristor and the charge from only this memristor was considered (as shown in the flowchart in Fig. 6). Hence, the sneak path problem is not present in our method of implementing ICA in the crossbar.

In order to ascertain the robustness of the novel memristor-based implementation of ACY ICA, and Fast ICA, we demonstrated the framework on images consisting of mixed faces and evaluated its efficacy in obtaining the ICs from the mixed pixel image. The results of ICA were simulated on MATLAB software, 2015 version in order to determine the percentage of improvement in the image quality. Two original images each of size 512 × 512 pixels are reshaped to a one-dimensional signal of dimensions 1 × 262,144. The two original images are mixed to get two input images to obtain the appropriate input images to the memristor crossbar arrays. The first mixed image is fed to row 1 of the crossbar array (Fig. 3), of size 2 × 2, while the second mixed image is sent to the second row of the crossbar array. Pulse width modulation has been employed to store the intensity values of each pixel of the input images. The output charges obtained from the columns of the crossbar array using (10) is stored and fed to (13), so as to obtain the output. However, in order to update the memristor weights, the duration of a write pulse is required, which can be derived from (16).

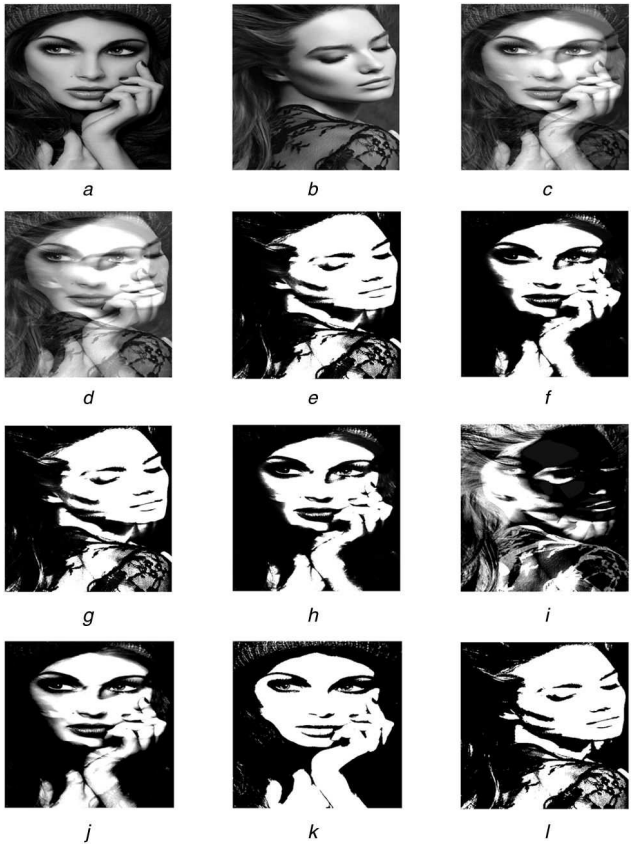

**Fig. 7** *Results of separating mixed images using the memristor-based hardware implementation of ICA*
*(a)*, *(b)* Original images, *(c)*, *(d)* Mixed images that are the inputs to the algorithm, *(e)*, *(f)* Software implementation outputs from ACY ICA, *(g)*, *(h)* Memristive crossbar array outputs of ACY ICA, *(i)*, *(j)* Software implementation outputs of Fast ICA, *(k)*, *(l)* Memristive implementation of Fast ICA

**Table 1** Percentage of improvement of quality measures for memristor-based architecture with respect to the simulation-based image processing results

| Percentage of improvement of quality measures, % | Memristive ACY ICA | Memristive fast ICA |
|---|---|---|
| percentage in SSIM | 0.03 | 67.27 |
| percentage in GSM | 0.01 | 3.22 |
| percentage in PSNR | 0.09 | 36.95 |
| percentage in MSE | 0.25 | 61.88 |

**Table 2** Percentage of improvement of quality measures for memristor-based architecture with respect to the simulation-based image processing results including device variations

| Percentage of improvement of quality measures, % | Memristive ACY ICA with 3% variations | Memristive Fast ICA with 3% variations |
|---|---|---|
| percentage in SSIM | −0.17 | 66.92 |
| percentage in GSM | −0.18 | 2.70 |
| percentage in PSNR | 0.02 | 35.84 |
| percentage in MSE | 0.16 | 60.78 |

The original images are depicted in Figs. 7*a* and *b*, whereas the mixed images that are the inputs to the algorithm are shown in Figs. 7*c* and *d*. Figs. 7*e* and *f* show the processed images obtained from the software implementation of ACY ICA. On the other side, the images obtained from the memristive crossbar-based ACY ICA are presented in Figs. 7*g* and *h*. Furthermore, the processed images obtained from the Fast ICA techniques are shown in Figs. 7*i* and *j*, and those obtained from the memristive crossbar-based Fast ICA are delineated in Figs. 7*k* and *l*.

The percentage of improvement in SSIM, GSM, PSNR and MSE values are tabulated in Table 1. It is evident that the contrast of both processed images is improved using memristor-based implementation of ACY ICA and Fast ICA algorithms. The percentage of improvement of SSIM qualities indices such as SSIM and GSM are high in proposed memristor-based architecture as compared to the software one.

In order to address the robustness of the system or investigate device-to-device variability in representing weight values, Monte Carlo simulation has been performed and the obtained results were tabulated in Table 2. The variations were accomplished by the standard Monte Carlo process, i.e. a Gaussian probability distribution of the variables and a standard deviation of 3% of the mean value was chosen for the device diameter (mean value 50 nm), high resistance state (mean value, 152 MΩ) and the low resistance state (mean value, 150 kΩ).

The effect of device variations on all four quality measures (SSIM, GSM, PSNR and MSE) are provided in Table 2. For Fast ICA's memristor implementation, the changes in parameters were negligible (only 1–2%) when compared with the ideal simulation results (without device variations). However, the Monte Carlo results were found to be better than conventional software implementation. On the other side, when the same were compared with ACY ICA's memristor implementation, the GSM and SSIM metrics are slightly degraded as compared to the conventional software implementation. The device variations may lead to different levels of polarisations within the memristor, which might offer inaccurate weight values to be stored in the device, thus the slight degradations only for ACY ICA were observed. In this regard, it is important to mention that our aim was to investigate the performance parameters for Fast ICA since this is the state-of-the-art ICA and much advantageous from its ACY ICA counterpart. Interestingly, the obtained results from Fast ICA were up to the mark even when the device-to-device variability was considered, which validates the robustness of the proposed system.

The synaptic or logic studies on memristor were celebrated widely; however, very little attention has been paid to implement this ICA work by utilising memristors. For real-time processing of images, the FPGA-based architecture is normally used for the implementation of the ICA algorithm [21–23]. However, the FPGA is slower, and requires more power to operate and offers high propagation delay which may not be suitable for real-time image source separation applications [21]. Apart from this, ASICs are considered as an alternative approach to implement ICA [24, 25]. However, both the FPGA and ASICs designs are developed upon CMOS devices, which are limited by the Von-Neumann bottleneck. Moreover, they were capable of implementing only the given specific algorithm. When the algorithm gets changed, the entire design has to be changed. On the other side, the memristive crossbar can implement any machine-learning algorithm, which is based on multiply and accumulate functions [35]. In addition, in this work, the proposed memristor-based crossbar architecture has better switching performance, low power consumption and high-density integration, which are very helpful to store image information within the device. The other entropy-based ICA techniques and other unsupervised learning methods such as linear discriminant analysis; *K*-means clustering can also be implemented using the proposed memristor-based crossbar network.

## 6 Conclusion

A novel hardware implementation of the ICA algorithm was proposed using an innovative memristor crossbar array. In this work, the synaptic characteristics of Pt/Cu:ZnO/Nb:STO memristor was observed and this concept was employed to implement ACY ICA and Fast ICA neural networks. When images separation application is concerned, it was observed that memristive implementation is much superior to the conventional software-based approaches. The improvement in the case of the traditional ACY ICA is <1%, while in case of the state of the art entropy-based Fast ICA, it is about 67% when the memristive network was engaged. It was found that the proposed non-traditional is indeed effective, efficient, and contemporary, and it can be a suitable alternative to the conventional CMOS-based systems to implement advanced ICA. Another intriguing factor is the sneak paths, which may deteriorate the overall performance of a crossbar network. It is noteworthy to mention that in this work, the demonstrated memristive array is very small (just $2 \times 2$), thus the sneak paths current is not very significant. However, for larger networks, one may need to study the efficacy of sneak paths on the system's performance. This work paves the way forward for futuristic neural network systems for not only image separations applications, but also for many other crucial applications such as fingerprints separation, palm print recognition, and image de-noising.

## 7 Acknowledgments

All the authors sincerely acknowledge the financial support from BRNS, DAE, Govt. of India through project no. 34/14/11/2017-BRNS/34286 to accomplish this research work. P.K.R.B. is thankful to BRNS for supporting his fellowship for PhD. All the authors are also thankful to BITS-Pilani Hyderabad Campus and its cleanroom facilities for some technical supports. S.K. and S.B. also acknowledge the support from Dr D. Bhattacharya, BARC-Mumbai, who is also the PC for the project. P.K.R.B. and V.J.L. contributed equally to this work.

## 8 References

[1] Fouda, M.E., Neftci, E., Eltawil, A., *et al.*: 'Independent component analysis using RRAMs', *IEEE Trans. Nanotechnol.*, 2018, **18**, pp. 611–615
[2] Hyvärinen, A., Oja, E.: 'Independent component analysis: algorithms and applications', *Neural Netw.*, 2000, **13**, (4–5), pp. 411–430
[3] Oja, E.: 'Applications of independent component analysis'. Lecture Notes in Computer Science (including Subseries Lecture Notes in Artifical Intelligence and Lecture Notes in Bioinformatics), Berlin, 2004, (LNCS, 3316), pp. 1044–1051
[4] Haykin, S., Chen, Z.: 'The cocktail party problem', *Neural Comput.*, 2005, **17**, pp. 1875–1902
[5] Bell, A.J., Sejnowski, T.J.: 'An information-maximization approach to blind separation and blind deconvolution', *Neural Comput.*, 1995, **7**, (6), pp. 1129–1159
[6] Amari, S., Cichocki, A., Yang, H.: 'A new learning algorithm for blind signal separation', *Adv. Neural Inf. Process. Syst.*, 1996, **8**, pp. 757–763
[7] Jutten, C., Herault, J.: 'Blind separation of sources, part I: an adaptive algorithm based on neuromimetic architecture', *Signal Process.*, 1991, **24**, (1), pp. 1–10
[8] Roh, T., Hong, S., Cho, H., *et al.*: 'A 259.6 μW HRV-EEG processor with nonlinear chaotic analysis during mental tasks', *IEEE Trans. Biomed. Circuits Syst.*, 2016, **10**, (1), pp. 209–218
[9] Shih, W.-Y., Liao, J.-C., Huang, K.-J., *et al.*: 'An efficient VLSI implementation of on-line recursive ICA processor for real-time multi-channel EEG signal separation'. 35th Annual Int. Conf. of the IEEE Engineering in Medicine and Biology Society (EMBC), Osaka, Japan, 2013, pp. 6808–6811
[10] Vourkas, I., Sirakoulis, G.C.: 'Emerging memristor-based logic circuit design approaches: a review', *IEEE Circuits Syst. Mag.*, 2016, **16**, (3), pp. 15–30
[11] Choi, S., Sheridan, P., Lu, W.D.: 'Data clustering using memristor networks', *Sci. Reports*, 2015, **5**, p. 10492
[12] Hamdioui, S., Kvatinsky, S., Cauwenberghs, G., *et al.*: 'Memristor for computing: myth or reality?'. Proc. Design, Automation and Test in Europe, Lausanne, Switzerland, 2017, pp. 722–731
[13] Jeong, H., Shi, L.: 'Memristor devices for neural networks', *J. Phys. D: Appl. Phys.*, 2018, **52**, (2), pp. 023003
[14] Choi, S., Shin, J.H., Lee, J., *et al.*: 'Experimental demonstration of feature extraction and dimensionality reduction using memristor networks', *Nano Lett.*, 2017, **17**, (5), pp. 3113–3118
[15] Sheridan, P.M., Cai, F., Du, C., *et al.*: 'Sparse coding with memristor networks', *Nature Nanotechnol.*, 2017, **12**, (8), p. 784
[16] Haj-Ali, A., Ben-Hur, R., Wald, N., *et al.*: 'IMAGING: in-memory algorithms for image processing', *IEEE Trans. Circuits Syst. I: Regular Papers*, 2018, **65**, (12), pp. 4258–4271
[17] Rák, A., Cserey, G.: 'Independent component analysis by memristor based neural networks'. 14th Int. Workshop on Cellular Nanoscale Networks and their Applications, Notre Dame, USA, 2014, pp. 1–2
[18] Mazady, A., Anwar, M.: 'Memristor: part I – the underlying physics and conduction mechanism', *IEEE Trans. Electron. Devices*, 2014, **61**, (4), pp. 1054–1061
[19] Hyvarinen, A.: 'Fast and robust fixed-point algorithms for independent component analysis', *IEEE Trans. Neural Netw.*, 1999, **10**, (3), pp. 626–634
[20] Boukouvalas, Z., Mowakeaa, R., Fu, G.-S., *et al.*: 'Independent component analysis by entropy maximization with Kernels', 2016, **1**, pp. 1–6, ArXiv e-prints, October 2016, available: https://arxiv.org/pdf/1610.07104.pdf
[21] Charoensak, C., Sattar, F.: 'Design of low-cost FPGA hardware for real-time ICA-based blind source separation algorithm', *Eurasip J. Appl. Signal Process.*, 2005, **173453**, (18), pp. 3076–3086, available: https://doi.org/10.1155/ASP.2005.3076
[22] Wang, Y., Lin, Q.: 'FPGA implementation of one-unit fixed-point ICA-R algorithm'. Proc. Int. Conf. on Intelligent Control and Information Processing, ICICIP, Dalian, China, 2010, pp. 191–194
[23] Morales, D.P., García, A., Castillo, E., *et al.*: 'An application of reconfigurable technologies for non-invasive fetal heart rate extraction', *Med. Eng. Phys.*, 2013, **35**, (7), pp. 1005–1014
[24] Stanacevic, M., Li, S., Cauwenberghs, G.: 'Micropower mixed-signal VLSI independent component analysis for gradient flow acoustic source separation', *IEEE Trans. Circuits Syst. I: Regular Papers*, 2016, **63**, (7), pp. 972–981
[25] Fang, W.C., Huang, K.J., Chou, C.C., *et al.*: 'An efficient ASIC implementation of 16-channel on-line recursive ICA processor for real-time EEG system'. 36th Annual Int. Conf. of the IEEE Engineering in Medicine and Biology Society, EMBC, Chicago, Illinois, 2014, pp. 3849–3852
[26] Giannakopoulos, X., Karhunen, J., Oja, E.: 'An experimental comparison of neural ICA algorithms', *Int. J. Neural Syst.*, 1999, **9**, pp. 99–114
[27] Shamsi, J., Amirsoleimani, A., Mirzakuchaki, S., *et al.*: 'Modular neuron comprises of memristor-based synapse', *Neural Comput Appl*, 2017, **28**, (1), pp. 1–11
[28] Boppidi, P.K.R., Raj, P.M.P., Challagulla, S., *et al.*: 'Unveiling the dual role of chemically synthesized copper doped zinc oxide for resistive switching applications', *J. Appl. Phys.*, 2018, **124**, (21), p. 214901
[29] Mitchell, T.: 'Face images directory'. Available at http://www.cs.cmu.edu/~tom/faces.html, accessed May 2019
[30] Jo, S.H., Chang, T., Ebong, I., *et al.*: 'Nanoscale memristor device as synapse in neuromorphic systems', *Nano Lett.*, 2010, **10**, (4), pp. 1297–1301
[31] Suresh, B., Boppidi, P.K.R., Prabhakar Rao, B.V.V.S.N., *et al.*: 'Realizing spike-timing dependent plasticity learning rule in Pt/Cu:Zno/Nb:STO memristors for implementing single spike based denosing autoencoder', *J. Micromech. Microeng.*, 2019, **29**, (8), p. 085006
[32] Kvatinsky, S., Ramadan, M., Friedman, E.G., *et al.*: 'VTEAM: a general model for voltage-controlled memristors', *IEEE Trans. Circuits Syst. II: Express Briefs*, 2015, **62**, (8), pp. 786–790
[33] Yan, X., Wang, J., Zhao, M., *et al.*: 'Artificial electronic synapse characteristics of a Ta/Ta2O5-x/Al2O3/InGaZnO4 memristor device on flexible stainless steel substrate', *Appl. Phys. Lett.*, 2018, **113**, (1), p. 013503
[34] Zhou, Z., Yan, X., Zhao, J., *et al.*: 'Synapse behavior characterization and physical mechanism of a Tin/Sio x /p-Si tunneling memristor device', *J. Mater. Chem. C*, 2019, **7**, (6), pp. 1561–1567
[35] Liao, Y., Wu, H., Wan, W., *et al.*: 'Novel in-memory matrix-matrix multiplication with resistive cross-point arrays'. Digest of Technical Papers-Symp. on VLSI Technology, Honolulu, Hawaii, 2018, pp. 31–32